\documentclass[aps,twocolumn,groupedaddress]{revtex4}

\usepackage{graphics}
\usepackage{graphicx}

\def\fun#1#2{\lower3.6pt\vbox{\baselineskip0pt\lineskip.9pt
\ialign{$\mathsurround=0pt#1\hfil##\hfil$\crcr#2\crcr\sim\crcr}}}

\begin{document}

\title{Large Logarithms in the Beam Normal Spin Asymmetry \\
of Elastic Electron--Proton Scattering}

\date{\today}
\author{Andrei V. Afanasev$^{a)}$ and N.P. Merenkov$^{b)}$}
\affiliation{
$^{(a)}$Jefferson Lab, Newport News, VA 23606, USA\\
$^{(b)}$ NSC ''Kharkov Institute of Physics and Technology'',\\
Kharkov 61108, Ukraine}

\begin{abstract}
We study a parity-conserving single-spin beam asymmetry of elastic electron-proton scattering
induced by an absorptive part of the two-photon exchange amplitude. It is demonstrated that
excitation of inelastic hadronic intermediate states by the consecutive exchange of two photons
leads to logarithmic and double-logarithmic enhancement due to contributions of hard
collinear quasi-real photons. The asymmetry at small electron scattering
angles is expressed in terms of the total photoproduction cross section on the proton,
and is predicted to reach the magnitude of 20-30 parts per million.
At these conditions and fixed 4-momentum transfers, the asymmetry
is rising logarithmically with increasing electron beam energy, following the
high-energy diffractive behavior of total photoproduction cross section on the proton.

\end{abstract}

\maketitle

\section{Introduction}

Recently the two--photon exchange (TPE) mechanism in elastic
electron--proton scattering started to draw a lot of attention.
The reason is that this mechanism possibly accounts for the
difference between the high--$Q^2$ values of the ratio
$G_{Ep}/G_{Mp}$ \cite{Mel} measured in unpolarized and polarized
electron scattering. Calculations of Ref.\cite{Afanas} using a
formalism of Generalized Parton Distributions \cite{GPD} confirm
such a possibility and decisive experimental tests are being
proposed \cite{Brooks}.

On the other hand, it has been known for a long time \cite{ST1,Barut,Ru} that the TPE mechanism can
generate the single-spin normal asymmetry (SSNA) of electron scattering due to a nonzero imaginary
($\Im$) part of the TPE amplitude $A_{2\gamma}$,
 \begin{equation}\label{1}
 A_n=\frac{2A_{Born}\Im (A^*_{2\gamma})}{|A_{Born}|^2} \,
 \end{equation}
where the one-photon-exchange amplitude $A_{Born}$ is purely real due to time-reversal
invariance of electromagnetic interactions.

Our earlier calculations of the TPE effect on the proton
\cite{aam} predicted the magnitude of beam SSNA at the level of a
few parts per million (ppm). The effect appears to be small due to
two suppression factors combined: $\alpha=1/137$, since the effect
is higher-order in the electromagnetic interaction; and a factor
of electron mass $m_e$ arising due to electron helicity flip. The
predictions of Ref.\cite{aam}, assuming no inelastic excitations
of the intermediate proton, used only proton elastic form factors
as input parameters and appeared to be in qualitative agreement
with experimental data from MIT/Bates \cite{Wells}. The result of
Ref.\cite{aam} with an elastic intermediate proton state was
reproduced later in Ref.\cite{Mark}. In another calculation of
beam SSNA of Ref.\cite{Musolf}, a low-momentum expansion was used
for the TPE loop integral, which resulted in approximate analytic
expressions valid for low electron beam energies. The main
theoretical problem in description of the TPE amplitude on the
proton at higher energies in the GeV range is a large uncertainty
in the contribution of the inelastic hadronic intermediate states.
In Ref.\cite{Mark} the beam SSNA at large momentum transfers was
estimated at the level of one ppm, using the partonic framework
developed in Ref.\cite{Afanas} for TPE effects not related to the
electron helicity flip.

Current experiments designed for parity-violating
electron scattering allow to measure the beam asymmetry with a
fraction of ppm accuracy \cite{Maas, E158,PViol} and may also
provide data on the parity-conserving beam SSNA. In fact, such
measurements are needed because beam SSNA is a source of
systematic corrections in the measurements of parity-violating
observables.

During our previous work \cite{aam} we noticed that while considering
excitation of inelastic intermediate hadronic states, the expressions for the
beam SSNA (Eq.(11) of Ref.\cite{aam}), after factoring out the electron mass,
have an enhancement when at least one of the photons in the
TPE loop integral is collinear to its parent electron, i.e. the virtuality
of the exchanged photon is of the order of electron mass. It is interesting that
this effect did not appear for the target SSNA calculations \cite{aam}.
Independently, enhancement due to exchange of collinear photons
for the beam SSNA was observed by other authors
Ref.\cite{Pasquini}, who considered the hadronic intermediate states in the TPE amplitude
in the nucleon resonance region using a phenomenological model (MAID)
for single--pion electroproduction.

In this paper, we study the analytic structure of the
collinear-photon exchange in the TPE amplitude and demonstrate
that it results in enhancement of the beam SSNA described by
single and double logarithms of the type $\log(Q^2/m_e^2)$ and
$\log^2(Q^2/m_e^2)$. Using a general requirement of gauge
invariance of the nonforward nucleon Compton tensor, we find that
such enhancement does not take place for the target SSNA (with
unpolarized electrons) and spin correlations caused by
longitudinal polarization of the scattering electrons.  When
modelling the TPE mechanism with nucleon resonance excitation, we
observe that depending on the electron beam energy, the beam SSNA
has a double-logarithmic enhancement if the Mandelstam variable
$\sqrt{s}$ nears the resonance mass, and a single-logarithmic
enhancement otherwise. For electron energies above the resonance
region and small scattering angles, we use an optical theorem to
relate the nucleon Compton amplitude to the total photoproduction
cross section and obtain a simple analytic formula for the beam
SSNA in this kinematics.


\section{Properties of leptonic tensor}
\begin{figure}
\includegraphics[width=7cm]{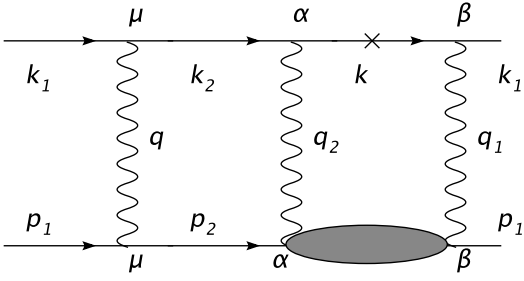}
\caption{Interference between the Born and the TPE box diagrams in elastic
e-p scattering that determined SSNA.}
\label{diagrams}
\end{figure}

First, we write the formula for SSNA in terms of rank-3 leptonic
and hadronic tensors which appear in the interference between the Born and TPE amplitudes
as shown in Fig.\ref{diagrams}.

\begin{equation}\label{2}
A_n=\frac{-i\alpha
Q^2}{\pi^2D(s,Q^2)}\int\frac{d^3k}{2E_k}\frac{L_{\mu\alpha\beta}
H_{\mu\alpha\beta}}{q_1^2 q_2^2} \ ,
\end{equation}
where $Q^2=-q^2$, $k (E_k)$ is the 3-momentum (energy) of the
intermediate on-mass-shell electron in the TPE box diagram, $q_1$ and
$q_2$ are the 4-momenta of the intermediate photons, $q_1-q_2=q$.
The factor $Q^2/D(s,Q^2)$ in Eq.(\ref{2})
is due to the squared Born amplitude, namely,

\begin{equation}\label{3}
D(s,Q^2) = \frac{Q^4}{2}\big(F_1+F_2\big)^2 +[(s-M^2)^2-
\end{equation}
$$Q^2s]\Big(F_1^2+\frac{Q^2}{4M^2} F_2^2\Big ),$$ where $F_1(F_2)$
is the Dirac (Pauli) proton form factor, $M$ is the proton mass
and $s=(k_1+p_1)^2$ is a Mandelstam variable. Our sign convention
for the beam asymmetry follows from the definition of the normal
vector with respect to the electron scattering plane:
${\bf{k}_1}\times {\bf{k}_2}.$

Using the above notation, we have
\begin{equation}\label{4}
L_{\mu\alpha\beta}=\frac{1}{4}Tr(\hat{k}_2+m_e)\gamma_{\mu}(\hat{k}_1+m_e)(1-
\gamma_5\hat{\xi^e})\gamma_{\beta}(\hat{k}+m_e)\gamma_{\alpha}\ ,
\end{equation}
and
\begin{equation}\label{5}
H_{\mu\alpha\beta}=\frac{1}{4}Tr(\hat{p}_2+M)\Gamma_{\mu}(\hat{p}_1+M)(1-
\gamma_5\hat{\xi^p})\Im T_{\beta\alpha}\ ,
\end{equation}
where $m_e$ is the electron mass, $\xi^e(\xi^p)$ is the
polarization 4-vector of the electron beam (proton target),
$\Gamma_{\mu}=\gamma_{\mu}(F_1+F_2)-(p_{1\mu}+p_{2\mu})F_2/(2M)$,
and $T_{\beta\alpha}$ is in general  a non-forward proton Compton
tensor that describes any possible hadronic intermediate states in the TPE amplitude, while
the symbol $\Im$ denotes the imaginary
(absorptive) part. Thus, we see that in accordance with Eq.(2) the
single-spin normal asymmetry probes the imaginary part of
contraction of the leptonic and hadronic tensors defined by
Eqs.(\ref{4}) and (\ref{5}), respectively. In turn, this imaginary
part is related with the imaginary part of the nucleon non-forward
Compton tensor $\Im T_{\beta\alpha}$. It was noted by De Rujula {\it et al.}
\cite{Ru} awhile ago for the case of normal polarization of the proton target.
For the case of beam SSNA on a proton, the first calculation was done in Ref.\cite{aam}.

After some algebra we arrive at the following expression for the model-independent
leptonic tensor
\begin{equation}\label{6}
L_{\mu\alpha\beta}=L^{(un)}_{\mu\alpha\beta} +
L^{(pol)}_{\mu\alpha\beta} \ ,
\end{equation}
where the spin--independent part is

$$L^{(un)}_{\mu\alpha\beta}=
\frac{1}{2}q_1^2(g_{\mu\beta}k_{2\alpha}-g_{\mu\alpha}k_{2\beta})-
\frac{1}{2}q_2^2(g_{\mu\beta}k_{1\alpha}-g_{\mu\alpha}k_{1\beta})-$$
\begin{equation}\label{7}
 k_{\mu}[k_1k_2]_{\alpha\beta}+
\frac{1}{2}g_{\alpha\beta}(q_1^2k_{2\mu}+q_2^2k_{1\mu}-q^2k_{\mu})+
\end{equation}
$$ \frac{1}{2}q^2 (g_{\mu\alpha}k_{\beta}+g_{\mu\beta}k_{\alpha})
+k_{2\mu}(kk_1)_{\alpha\beta} +k_{1\mu}(kk_2)_{\alpha\beta}\ ,$$
$$[ab]_{\alpha\beta} = a_{\alpha}b_{\beta} - a_{\beta}b_{\alpha}\
, \ \ (ab)_{\alpha\beta} = a_{\alpha}b_{\beta} +
a_{\beta}b_{\alpha}\ ,$$ and the spin--dependent part is given by
$$L^{(pol)}_{\mu\alpha\beta}=im_e\bigg[-g_{\alpha\beta}(\mu q q_2
\xi^e) + k_{\beta}(\mu\alpha q \xi^e)+k_{\alpha}(\mu\beta q
\xi^e)$$
\begin{equation}\label{8}
+\xi^e_{\mu}(\alpha\beta q q_2) + (\xi^ek_2)(\mu\alpha\beta q_1) +
k_{2\mu}(\alpha\beta q_1 \xi^e)+
\end{equation}
$$ k_{1\mu}(\alpha\beta q_2 \xi^e)+ \frac{1}{2}q^2(\mu\alpha\beta
\xi^e)\bigg]\ ,$$
$$(abcd)\equiv\epsilon_{\nu\lambda\rho\sigma}a_{\nu}b_{\lambda}c_{\rho}
d_{\sigma}\ , $$
where the on-shell condition $k^2_{\mu}=m_e^2$ was used for
the intermediate electron 4-momentum.
The above  leptonic and hadronic tensors
satisfy the conditions $$L_{\mu\alpha\beta}q_{\mu} =
L_{\mu\alpha\beta}q_{2\alpha} = L_{\mu\alpha\beta}q_{1\beta} = 0$$
\begin{equation}\label{9}
H_{\mu\alpha\beta}q_{\mu} = H_{\mu\alpha\beta}q_{2\alpha} =
H_{\mu\alpha\beta}q_{1\beta} = 0
\end{equation}
separately for spin--independent and spin--dependent parts, as follows from
gauge invariance of electromagnetic interactions.

Let us consider the leptonic tensor
in the limiting case when one of the intermediate photon in the box diagram is
collinear to its parent electron, for example, when $q_1=xk_1, \ x\simeq 1.$
Note first that for elastic proton intermediate state such kinematics is not allowed
by the 4-momentum conservation,  $(xk_1+p_1)^2=M^2$. But
for the inelastic intermediate excitations we have $x=(W^2-M^2)/(s-M^2),$ where $W^2$ is the squared
invariant mass of the intermediate hadronic system.

It can be seen, using the relations
\begin{equation}\label{10}
q_2^2=(1-x)q^2, \ \ k=(1-x)k_1, \ \ q_1^2\simeq 0
\end{equation}
that in the considered conditions the unpolarized leptonic tensor is given by
\begin{equation}\label{11}
L^{(un)}_{\mu\alpha\beta}=\frac{1-x}{x}q_{1\beta}
\Big[q^2g_{\mu\alpha}+2(k_1k_2)_{\mu\alpha}\Big]\ .
\end{equation}
Because any gauge-invariant hadronic tensor has to give zero after contracting with
$q_{1\beta}$ (see Eq.(\ref{9})), we conclude that collinear photon kinematics does
not contribute to the target SSNA (or recoil proton polarization)
which are defined by the spin--independent part of the
leptonic tensor $L_{\mu\alpha\beta}.$ This conclusion confirms
our previous calculations \cite{aam}.

For the spin--dependent part of leptonic tensor, we obtain in the
considered limit,
$$\frac{1}{im_e}L^{(pol)}_{\mu\alpha\beta}\Big|_{q_1\to
xk_1}=\frac{2(1-x)}{x}q_{1\beta}(\mu\alpha q \xi^e)+$$
\begin{equation}\label{eq:12}
x\Big[-g_{\alpha\beta}(\mu q k_1 \xi^e)
+\frac{q^2}{2}(\mu\alpha\beta \xi^e)+\xi^e_{\mu}(\alpha\beta q
k_1)+
\end{equation}
$$(\xi^ek_2)(\mu\alpha\beta k_1) + k_{2\mu}(\alpha\beta k_1 \xi^e)
+ k_{1\mu}(\alpha\beta k_2 \xi^e) \Big]\ . $$
If the electron beam
is polarized longitudinally, ($\xi^e_{\mu} \simeq k_{1\mu}/m_e$),
the term in the square brackets of Eq.(\ref{eq:12}) turns to zero, and we have the same
situation as in the case of unpolarized beam, namely,  the region of the
small $q_1^2$ (or $q_2^2$) does not contribute when the
intermediate photon is collinear to its parent
electron.

A different phenomenon takes place in the case of the normal polarized electron
beam
\begin{equation}\label{13}
\xi^e_{\mu}=\frac{2(\mu k_1 p_1 q)}{\sqrt{Q^2[(s-M^2)^2-Q^2s]}}\ .
\end{equation}
In this case the term in square brackets of Eq.(\ref{eq:12}) is not zero and the considered collinear photon
kinematics contributes
with essential logarithmic enhancement. Moreover, here we will demonstrate using specific examples that
this enhancement can be double-logarithmic.

Therefore, conservation of the electromagnetic
current that follows from gauge invariance (Eq.(\ref{9}))
is the reason why the collinear intermediate photons appear in the TPE contribution to the
beam SSNA, but not to the target SSNA. By analogy, we do not anticipate
contributions from collinear-photon exchange in unpolarized electron-proton
scattering, parity-violating asymmetries due to longitudinal electron polarization,
charged current neutrino-nucleon scattering and/or lepton weak capture if
the normal polarization of leptons is not involved.

\section{Hadronic tensor in the resonance region}

We first study the analytic properties TPE loop integration in a resonance
model taking into account nucleon resonances with quantum numbers $N^*\Big(\frac{1^+}{2}\Big), \
N^*\Big(\frac{1^-}{2}\Big)$ and $\Delta\Big(\frac{3^+}{2}\Big)$ as intermediate hadronic
states. In this model the imaginary part of the nucleon Compton tensor is given by
\begin{equation}\label{14}
\Im T_{\beta\alpha} = \Im\Big[T_{\beta\alpha}(+) + T_{\beta\alpha}(-) +
T_{\beta\alpha}(\Delta) \Big] \ .
\end{equation}

Using a general Lorentz structure of nucleon resonance excitations,
we verified that the contribution from collinear kinematics
of intermediate photons is zero for the quantity $L_{\mu\alpha\beta}
H_{\mu\alpha\beta}$ (Eq.(\ref{2})) for the unpolarized electron beam.
On the contrary, if the electron beam has normal polarization, the resulting expression
is proportional to the difference between the
resonance and the proton masses. Thus, we conclude that the collinear
intermediate photons can give a large contribution to the beam SSNA
in the resonance region.

For example, the general form of the integrand $L_{\mu\alpha\beta}
H_{\mu\alpha\beta}$ in the case of intermediate
Roper $N\Big(\frac{1^+}{2}\Big)$ resonance excitation is proportional to
\begin{equation}\label{15}
\Big(a+bq_1^2+cq_2^2+dq_1^2q_2^2 +eq_1^4+fq_2^4\Big)F_R(q_1^2)F_R(q_2^2) \ ,
\end{equation}
where coefficients $a,\ b, ...$ depend on $M,\ M_R, \ s$ and $q^2$ and $F_R(q_{1,2}^2)$
are the transition form factors of the resonance excitation.

The 3-dimensional loop integration in Eq.({\ref{2}) is done over all allowed angles of the
intermediate electron and the invariant mass of the intermediate hadronic state
in the range $M+m_{\pi}<W<\sqrt{s}-m_e$, as follows from the energy-momentum conservation, where
$W^2=(q_1+p_1)^2 = (q_2+p_2)^2$ and $m_\pi$ is a pion mass. Namely,
$$\frac{d^3k}{2E_k}= \frac{k\ dW^2}{4\sqrt{s}}d\Omega_k \ .$$
 In the c.m.s.
$$q_1^2=2m_e^2+2kk_1\cos{\theta_1}-2E_kE_{k1},$$
\begin{equation}\label{16}
q_2^2=2m_e^2+2kk_1\cos{\theta_2}-2E_kE_{k1},
\end{equation}
where $k_1$ and $E_{k1}$ are the 3-momentum and the energy of the initial
electron, $\theta_1(\theta_2)$ is the angle between 3-momenta
of the intermediate and the initial (scattered) electrons.

Collinear photon kinematics corresponds to $\cos{\theta_1}\simeq
1$ (or $\cos{\theta_2}\simeq 1$ when the quantities $q_1^2$ and
$q_2^2$ become small and change from their minimal value
$(2m_e^2+2kk_1-2E_kE_{k1})$ up to $E_kE_{k1}\theta_1^2.$ The most
singular term in the TPE diagram integral at such conditions comes
from the coefficient $a$ in Eq.(\ref{15}). We use a substraction procedure to
present the
result of angular integration of this term,
 $$\int\frac{d\Omega_k}{q_1^2q_2^2}F_R(q_1^2)F_R(q_2^2)=
\int\frac{d\Omega_k}{q_1^2q_2^2}\Phi_0(q_1^2,q_2^2) -$$
$$\frac{2\pi}{E_{k1}k}\int
\limits_{-1}^1\frac{dc_1}{|c-c_1|}\Big[\frac{\Phi_1(q_1^2)}{q_1^2}
-\frac{\Phi_1(q_{1c}^2)}
{q_{1c}^2}\Big]- $$
\begin{equation}\label{17}
\frac{2\pi}{q_{1c}^2E_{k1}k}\Phi_1(q_{1c}^2)L_2 +\frac{\pi}{Q^2k^2}F^2_R(0)L_1 \ ,
\end{equation}
where $c=\cos{\theta}$, with $\theta$ being the electron c.m.s.
scattering angle, $ c_1=\cos{\theta_1}$,
$q^2_{1c}=q_1^2(c_1\rightarrow c)$ and $$\Phi_0(q_1^2,q_2^2) =
F_R(q_1^2)F_R(q_2^2)-F_R(0)[F_R(q_1^2)+F_R(q_2^2)] +$$ $$F^2_R(0),
\ \ \ \Phi_1(q_1^2) = F_R(0)[F_R(q_1^2)-F_R(0)]\ , $$ $$
L_1=\frac{1}{K}\log\frac{2K+1}{2K-1}, \ \
K=\sqrt{\frac{1}{4}+\eta}, \ \ \eta=
\frac{m_e^2(E_{k1}-E_k)^2}{Q^2k^2} \ ,$$ $$ L_2 =
\log\frac{A_+A_-}{m_e^2(E_{k1}-E_k)^2\sin^2{\theta}}\ , \
A_{\pm}=k_1k \pm E_{k1}E_kc $$
\begin{equation}\label{18} +\sqrt{(k_1k \pm
E_{k1}E_kc)^2+m_e^2\sin^2{\theta}(E_{k1}-E_k)^2}.
\end{equation}

The first two terms in the r.h.s. of Eq.(\ref{17}) are regular and
they can be easily integrated numerically (after choosing specific
parametrization of $F_R(q_1^2)$), whereas the last two ones,
enhanced logarithmically, appear due to proximity to the dynamical
pole that arises from collinear photon kinematics, and increased
precision is required if this problem were solved entirely by
numerical integration.

We can use this subtraction procedure to extract the large--logarithm
contributions of the other terms in the r.h.s. of
Eq.(\ref{15}). But it should be noted that at small values of $q_1^2$,
for example, the following relations holds for the quantity $q_2^2\simeq q^2(s-W^2)/(s-M^2).$
Therefore,
every additional power of $q_2^2 $ in the numerator in this case gives
an additional factor of the order $q^2/s$ as compared with
a contribution from the term $a/(q_1^2 q_2^2).$ Such contributions
will give only small corrections for the case of small-angle (low-$q^2$)
electron scattering.

Now let us consider $W$-integration that brings an additional enhancement
from the region of small momenta $k$ of the intermediate electron,
that appears in the denominators of the last two terms of Eq.(\ref{17}).
Small values of $k$ correspond to the intermediate hadronic
system picking the entire energy provided by the external electron
beam, namely, $W^2\approx s$.
Calculating the resonance contribution to the TPE loop integral,
one may restrict integration over $W^2$ to the resonance region
$M_R^2-\Gamma_R M_R< W^2<M_R^2+\Gamma_R M_R$,
in Eq.(2). If $s>>M_R^2$, small values of $k$ are not reached
and no additional enhancement results from this
integration, therefore the resonance contribution at
large energies may be enhanced only by the first power of large
logarithm. Moreover, the structure of the expression (15) implies
that in this case the effect may be only of the order of the ratio $M_R^2/s$
because there is no contributions can compensate a large denominator $D(s,Q^2)$
coming
from Born normalization. In such conditions the contribution of the
resonances to the beam SSNA becomes negligible at the electron
beam energies high enough so that the upper limit in $W$ of TPE
loop integral extends above the resonance region.

On the other hand, we may describe suppression of the resonance contribution
away from the resonance peak
by the absorptive part of Breit--Wigner factor
$F_{BW}$ which reads
$$F_{BW}=
\frac{\Gamma_{R}M_{R}}{(W^2-M^2_{R})^2+\Gamma^2_{R}M^2_{R}}
\ , $$ where $\Gamma_{R}$ is
the total width of the resonance.

Loop integration with the above Breit-Wigner factor was done
numerically using an adaptive multi-dimensional integration
technique, and the result will be discussed in Section V. In the
meantime, we demonstrate how a double-logarithmic enhancement
appears at the level of analytical formulae. Let us formally
extend the integration region with respect to $W^2$ up to its
upper limit of $W=\sqrt{s}-m_e$ allowed by kinematics and consider
the $W^2$ dependence coming only from the integral phase space,
while neglecting other $W$-dependent factors. Then we can perform
analytic integration and the result reads
\begin{equation}\label{19}
\int\frac{kdW^2}{4\sqrt{s}}\frac{\pi}{Q^2k^2}L_1 =
\frac{\pi}{4Q^2}
\Big(\log^2\frac{Q^2}{m_e^2}+\frac{4\pi^2}{3}\Big) \ ,
\end{equation}
\begin{equation}\label{20}
\int\frac{k\ dW^2}{4\sqrt{s}}\frac{\pi}{Q_{1c}^2kE_{k1}}L_2 =
\frac{\pi}{4Q^2}
\Big[\frac{1}{2}\log^2\frac{4E_{k1}^2}{m_e^2}+\frac{\pi^2}{3}+
\end{equation}
$$\frac{2}{1+c}Li_2\Big(-\frac{1+c} {1-c}\Big)\Big]\ ,$$
where $Li_2(x)$ is a dilogarithm. Thus, we
see that inegration over $W^2$ results in an additional large
logarithmic factor due to contribution of the region where
$W^2\simeq s$, when the energy of the intermediate electron
becomes very small. For resonance excitation, such situation takes
place for the electron beam energy such as $\sqrt{s}\simeq M_R$
only. But for multi--particle hadronic states with a continuously
varying invariant mass it should always manifest itself.

\section{Master formula for the beam asymmetry}

As we noted before, the resonance contribution in the beam SSNA dies out beyond the resonance
region. But the many--particle intermediate states can contribute into the imaginary part of Compton
tensor in the box diagram at large values of $W^2$ near $s.$ We verify that at small
$Q^2$ this contribution is double-logarithmically enhanced due to collinear kinematics when the values
$q_1^2$ and $q_2^2$ are small. In the following we develop a realistic unitarity-based model for
description of the imaginary part of the Compton tensor for such kinematics.

Small values of $Q^2$ correspond to the forward limit of nucleon virtual Compton
amplitude. On the other hand, because $q_1^2$ and $q_2^2$ are also
small because of the collinear photon contributions, we can relate
the forward Compton amplitude to the total photoproduction cross
section by real photons.

A general form of the Compton tensor $T_{\beta\alpha}$ in terms
of 18 independent invariant amplitudes that are free from
kinematical singularities and zeros was derived in
Ref.\cite{Tarrach}. Among these amplitudes we  choose the ones
that contribute at the limit $q^2\to 0$ and $q_1^2\to0.$ It
automatically constrains virtuality of the second photon to
$q_2^2\to0$.  There is only one structure that contains the tensor
$g_{\alpha\beta}$ and does not die off under the considered
conditions. It reads \cite{Tarrach}
$$T_{\beta\alpha}=\Big[-(\bar{p}\bar{q})^2g_{\alpha\beta}-
(q_1q_2)\bar{p}_{\alpha}\bar{p}_{\beta}
+(\bar{p}\bar{q})(\bar{p}_{\beta}q_{1\alpha} +$$
\begin{equation}\label{21}
\bar{p}_{\alpha}q_{2\beta})\Big] A(q_1^2,q_2^2,q^2,W^2) ,
\end{equation}
$$\bar{p}=\frac{1}{2}(p_1+p_2), \ \ \bar{q} =
\frac{1}{2}(q_1+q_2)\ .$$ It can be verified that
$T_{\beta\alpha}$ defined by the above equation satisfies the conditions
$T_{\beta\alpha}q_{1\beta}=T_{\beta\alpha}q_{2\alpha}=0.$ Taking
into account that in accordance with Eq.(\ref{9}), the terms containing
$q_{1\beta}$ and $q_{2\alpha}$ do not contribute in the
contruction with the leptonic tensor, we can rewrite expression
into square brackets as
$$\Big[-g_{\alpha\beta}-\frac{(q_1q_2)}{(W^2-M^2-q_1q_2)^2}\Big(4p_{1\alpha}p_{1\beta}+
2(p_1q)_{\alpha\beta} +q_{\alpha}q_{\beta}\Big)-$$
\begin{equation}\label{22}
\frac{[p_1q]_{\alpha\beta}}{W^2-M^2-q_1q_2}
\Big](W^2-M^2-q_1q_2)^2\ .
\end{equation}

The normalization convention is chosen such that the imaginary part of the quantity
$(W^2-M^2-q_1q_2)^2A(W^2,q^2=0,q_1^2=q_2^2)$ is connected with the inelastic proton
structure function $W_1(W^2,q_1^2)$ by the following relation
\begin{equation}\label{23}
(W^2-M^2-q_1q_2)^2 \Im A(W^2,q^2=0,q_1^2=q_2^2)=
\end{equation}
$$\frac{\pi}{M}W_1(W^2,q_1^2)\ ,$$ and $W_1$, in turn, defines the
total photoproduction cross section \cite{Lev} as
\begin{equation}\label{24}
W_1(W^2,0)=\frac{W^2-M^2}{8\pi^2\alpha}\sigma^{\gamma p}_{tot}(W^2)\ .
\end{equation}
Keeping in mind that the main contribution to the beam SSNA arises
from collinear photon kinematics, in our further calculations we can use
$$\Im T_{\alpha\beta}=\Big[-g_{\alpha\beta}-\frac{q_1q_2}{(W^2-M^2-q_1q_2)^2}
\Big(4p_{1\alpha}p_{1\beta}+ 2(p_1q)_{\alpha\beta}$$
\begin{equation}\label{25}
+q_{\alpha}q_{\beta}\Big)-\frac{[p_1q]_{\alpha\beta}}{W^2-M^2-q_1q_2}
\Big]\frac{\pi}{M}W_1(W^2,q_1^2) .
\end{equation}
in Eq.(\ref{5}).

It may seem at first that in the limiting case of
very small $Q^2$ we can omit all terms proportional to $q$ in the
r.h.s. of Eq.(25), keeping only the term
$$\Big[-g_{\alpha\beta}-\frac{4(q_1q_2)}{(W^2-M^2-q_1q_2)^2}
p_{1\alpha}p_{1\beta}\Big] \frac{\pi}{M}W_1(W^2,q_1^2),$$ that at
$q^2=0$ satisfies automatically the Callan--Gross relation. But
such approximation is valid only for the symmetric part of
$T_{\alpha\beta}$ with respect to the indexes $\alpha$ and
$\beta.$ The reason is that the corresponding symmetric part of
the leptonic tensor (see Eq.(\ref{8})) contains the momentum
transfer $q$, and keeping it in the symmetric part of hadronic
tensor leads after contraction to additional small terms of the
order at least $Q^2/W^2.$ On the other hand, the antisymmetric
part of leptonic tensor contains terms which do not include the
momentum $q.$ Therefore, the antisymmetric part in Eq.(\ref{25})
has to be retained because it contributes at the same order with
respect to $Q^2/W^2.$ Note, however, that this antisymmetric part
of the hadronic tensor is not related to the polarized nucleon
structure functions, but it comes about as a consequence of the
gauge-invariant structure of Eq.(\ref{21}) even for a spinless
hadronic target.

Thus, in the considered limit the hadronic tensor defined in
general by Eq.(\ref{5}) can be written in the following form
$$H_{\mu\alpha\beta}=2\pi W_1\Big(F_1-\tau
F_2\Big)\Big(-g_{\alpha\beta}
-\frac{[p_1q]_{\alpha\beta}}{W^2-M^2-q_1q_2}\Big)\ ,$$
\begin{equation}\label{26}
 \tau =\frac{Q^2}{4M^2}\ .
\end{equation}
When deriving this expression, we also
omit the terms proportional to $p_{1\alpha}p_{1\beta}$ because they are suppressed
by  an additional power of $Q^2$ due to the factor of
$(q_1q_2)$. Using the relations
$$-g_{\alpha\beta}p_{1\mu}L^{(pol)}_{\mu\alpha\beta}=2im_e[(p_1 q
q_1 \xi^e)+(k_1 p_1 q \xi^e)],$$
$$-[p_1q]_{\alpha\beta}p_{1\mu}L^{(pol)}_{\mu\alpha\beta} =
im_e(u-s)(p_1 q q_1 \xi^e),$$
$s+q^2+u=2M^2$ , which are valid for the
normal beam polarization
$((\xi^ek_1)=(\xi^ek_2)=(\xi^ep_1)=(\xi^ep_2)=0)$, and the
explicit form of 4--vector $\xi^e$ given by Eq.(\ref{13}), we arrive at
\begin{equation}\label{27}
L^{(pol)}_{\mu\alpha\beta}H_{\mu\alpha\beta}=im_e\sqrt{Q^2}\Big(F_1-\tau
F_2\Big)\frac{(W^2-M^2)^2} {4\pi\alpha}\sigma_T(W^2,q_1^2)\ ,
\end{equation}
where $\sigma_T(W^2,q_1^2)$ is the total photoproduction cross section
with the transverse virtual photons. Now we combine this
expression with the formula (2) for the beam SSNA and perform
analytic integration. When integrating we take
$\sigma_T(W^2,q_1^2)\rightarrow\sigma^{\gamma p}_{tot}(W^2)$ and
assume  $\sigma^{\gamma p}_{tot}(W^2)$ to be constant with energy
($\approx$ 0.1 mb, according to Ref.\cite{pdg}).  The angular
integration results in a large logarithm $L_1$ defined in
Eq.(\ref{18}). Integration with respect to $W^2$ produces
double-logarithmic enhancement in the final result. In the
integration, special care needs to be taken of the region of small
energies of the intermediate electron, see Appendix for the
details. As a result, the master formula, that defines the beam
SSNA for small values of $Q^2$ and takes into account
contributions from collinear intermediate photons in the
TPE box diagram, has the following form
$$A_n^e=\frac{m_e\sqrt{Q^2}\sigma^{\gamma
p}_{tot}}{16\pi^2}\frac{F_1-\tau F_2}{F_1^2+\tau F_2^2}\times $$
\begin{equation}\label{28}
\Big(\log^2\frac{Q^2}{m_e^2}-
6\log\frac{Q^2}{m_e^2}+\frac{4\pi^2}{3} +4\Big)\ .
\end{equation}
One can see that at fixed values of $Q^2$ the beam SSNA does not
depend on the beam energy if the total photoproduction cross
section is energy-independent. This remarkable property of
small-angle beam SSNA follows from unitarity of the scattering
matrix and does not rely on a specific model of nucleon structure.

\section{Numerical Results and Discussion}
 First we analyze the general features of the beam SSNA in the nucleon resonance region.
 We perform 3-dimensional numerical integration in Eq.(\ref{2}), selecting the most
 singular term from Eq.(\ref{15}) in front of the coefficient $a$. The integral
 $$
 \frac{2 Q^2}{\pi}\int\frac{d^3k}{2E_k}\frac{F_{BW}(W)}{q_1^2 q_2^2}
 $$
 is shown in Fig.\ref{fig:rescont} using different assumptions about the energy-dependent integrand.
 For the plots of Fig.\ref{fig:rescont}, we fix $Q^2$= 0.05 GeV$^2$ and vary the electron beam
 energy. One can see that if we choose the mass and width of a $\Delta(1232)$-resonance,
 the result of the integration is still strongly peaked at the electron beam energy close
 to the position of this resonance in real photoproduction with same photon beam energies.
 The position of the $\Delta$-peak is slightly shifted to the higher energies (by about 35 MeV),
 which corresponds to the intermediate electron carrying a c.m.s. energy of about fifty electron masses.
 If the TPE integral were fully dominated by the region of small $k$, the result would be given by
 a dotted line in Fig.\ref{fig:rescont}.
 If the above integral is calculated  with an energy-independent
 nonresonant background
 which we take for illustrative purposes at 1/5th of the $\Delta$-resonance peak value,
 we see that the resonance contribution dies off at higher electron beam energies, confirming the
 analytic arguments of Section III. It can also be seen from Fig.\ref{fig:rescont} that
 the analytic formula of Eq.(\ref{19}) gives a good description of integration
 of energy-independent terms at the beam energies above the resonance region.

\begin{figure}
\includegraphics[width=7cm]{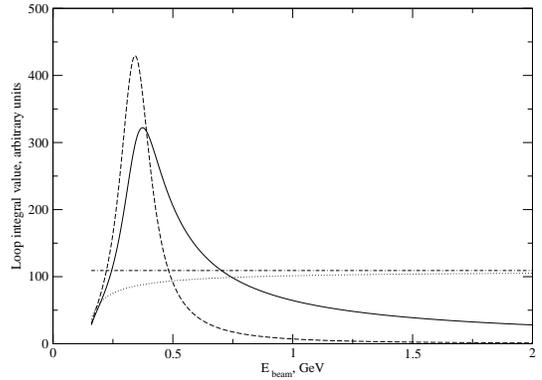}
\caption{Beam energy dependence of the loop integral (see the text for details) at fixed $Q^2=0.05 GeV^2$
using a Breit-Wigner factor $F_{BW}(W)$ with a) parameters of $\Delta$-resonance (solid line),
b) replacing $F_{BW}(W\to\sqrt{s})$ (dashed line), c) same as a) but replacing $F_{BW}(W)$ by 1/5th
of its peak value at W=M$_R$ (dotted line), and d) is same as c) but using an analytic formula of Eq.(\ref{19})
(dash-dotted line).
}
\label{fig:rescont}
\end{figure}

The master formula for beam SSNA Eq.(\ref{28}) neglects possible
$Q^2$ dependence of the invariant form factor of the nucleon
Compton amplitude, which was taken in its forward limit during the
derivation. In numerical calculations, we estimate additional
$Q^2$ (= Mandelstam $t$) dependence by introducing an empirical
form factor that was measured experimentally in the Compton
scattering on the nucleon in the diffractive regime (see
\cite{Bauer} for review). In the following, we use an exponential
suppression factor for the nucleon Compton amplitude $\exp{(-B
Q^2/2)}$, choosing the parameter $B$=8 GeV$^{-2}$ that gives a
good description of the nucleon Compton cross section from the
optical point to $-t\approx$ 0.8 GeV$^2$ (see Table V of
Ref.\cite{Bauer}). The predictions of Eq.(\ref{28}) combined with
the above described exponential suppression are presented in
Fig.\ref{fig:thdep} for the electron scattering kinematics
relevant for the E158 experiment at SLAC \cite{E158}.  We choose
fit 1 of Ref.\cite{block} for the total photoproduction cross
section in Eq.(\ref{28}). Exact numerical loop integration of
Eq.(\ref{2}) and the analytic results of Eq.(\ref{28}) agree with
each other with accuracy better than 1\%. Contributions from the
resonance region ($W^2<$ 4 GeV$^2$) were estimated at 10-20\% at
beam energies of 3 GeV, but rapidly decreasing below 1\% at higher
energies. We also tested sensitivity of our results to $q_{1,2}^2$
dependence of the electroproduction structure function $W_1$
(Eq.(\ref{23})), taking various empirical parameterizations for
it. We found no sensitivity for SLAC E158 kinematics and only
moderate sensitivity ($\approx$ 10\%) when we extend our
calculation to lower energies ($\approx$3 GeV) and higher
$Q^2\approx$0.5 GeV$^2$. For beam energies of 45 GeV, numerical
integration shows that more than 95\% (80\%) of the result for
beam SSNA comes from the upper 1/2 (3/4) part of the
$W^2$-integration range. Based on the results of numerical
analysis, we conclude that the formula (\ref{28}) gives a good
description of beam SSNA at small $Q^2$ and large $s$ above the
resonance region.

We also calculated the contribution of the elastic intermediate
proton state to the beam SSNA for high energies and small electron
scattering angles using the formalism of Ref.\cite{aam} and found
it to be highly suppressed compared to the inelastic excitations.
For the kinematics of SLAC E158 \cite{E158}, this suppression is a
few orders of magnitude due to different angular and energy
behavior of these contributions.

\begin{figure}
\vspace{0.1in}
\includegraphics[width=7cm]{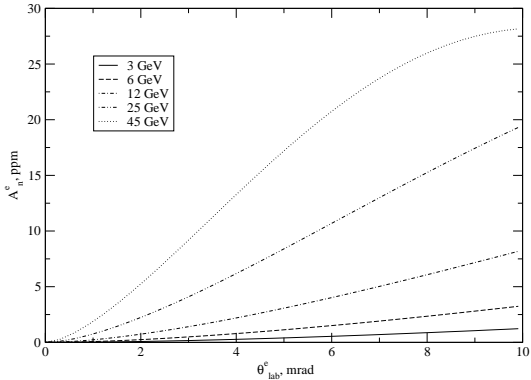}
\caption{Beam SSNA as a function of the lab scattering angle for different beam energies: 3 GeV (solid line),
6 GeV (dashed line), 12 GeV (dash-dotted line), 25 GeV (dash-double-dotted line) and 45 GeV (dotted line).}
\label{fig:thdep}
\end{figure}

Shown in Fig.\ref{fig:Q2dep} are the calculations for beam SSNA as
a function of $Q^2$ for different energies of incident electrons.
One can see that at small $Q^2$, the asymmetry follows
$\sqrt{Q^2}$ behavior described by Eq.(\ref{28}), while at higher
$Q^2$ the asymmetry turns over and starts to decrease due to the
introduced exponential form factor $\exp{(-B Q^2/2)}$. It can be
seen that at fixed $Q^2$ the magnitude of beam SSNA is predicted
to be approximately constant, as follows from slow logarithmic
energy dependence of the total photoproduction cross section.

\begin{figure}
\includegraphics[width=7cm]{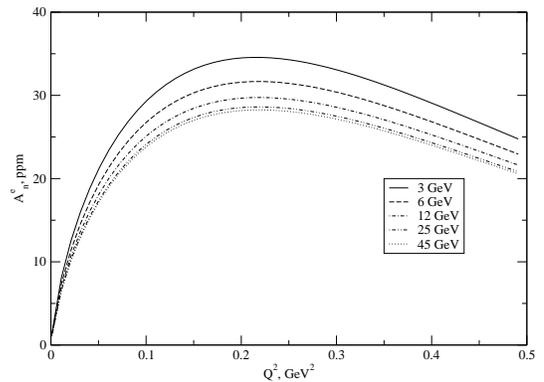}
\caption{Beam SSNA as a function of $Q^2$ for different beam energies. Notation is as in Fig.\ref{fig:thdep}.}
\label{fig:Q2dep}
\end{figure}

The latter feature is demonstrated in Fig.\ref{fig:sqsdep}, showing the calculated beam SSNA at fixed $Q^2$
in a wide energy range up to $\sqrt{s}$=500 GeV, where we used several parameterizations
for the total photoproduction cross section on a proton from Refs.\cite{block,DL}, shown in Fig.\ref{fig:sigtot}.
The physical reason for the almost constant photoproduction cross sections at high energies is believed to be soft
Pomeron exchange \cite{DL}, therefore the beam SSNA in the considered kinematics is sensitive to the physics
of soft diffraction.

The predicted $Q^2$ and energy dependence of beam SSNA, along with
its relatively large magnitude, is quite different from the model
expectations assuming that no hadronic intermediate states are
excited in the TPE amplitude.  Our unitarity-based model of
small-angle electron scattering predicts the magnitude of the beam
SSNA to reach 20-30~ppm in a wide range of beam energies. The good
news is that it makes beam SSNA measurable with a presently
reached fraction-of-ppm precision of parity-violating electron
scattering experiments \cite{PViol}. On the other hand, the
experiments measuring parity-violating observables need to use
special care to avoid possible systematic uncertainties due to the
parity-conserving beam SSNA. Fortunately, these effects can be
experimentally separated using different azimuthal dependence of
these asymmetries.

\begin{figure}
\includegraphics[width=7cm]{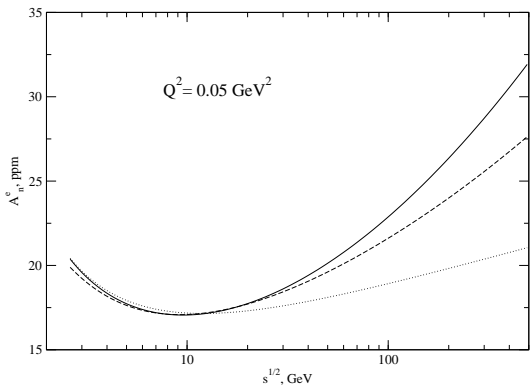}
\caption{Beam SSNA as a function of c.m.s. energy for fixed Q$^2$ = 0.05 GeV$^2$ for
different parameterizations of the total photoproduction cross section.
See Fig.\ref{fig:sigtot} for notation.}
\label{fig:sqsdep}
\end{figure}

\begin{figure}[h]
\vspace{0.2in}
\includegraphics[width=7cm]{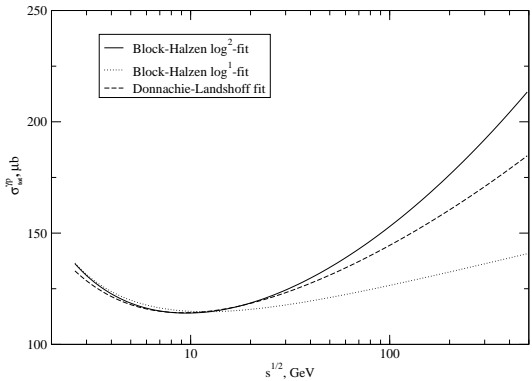}
\caption{Different parameterizations of total photoproduction cross section at high energies
used in the present calculation. A solid (dotted) line is a double-logarithmic fit 1
(single-logarithmic fit 3) from Block and Halzen
\cite{block}
and a dashed line is an original Donnachie and Landshoff fit \cite{DL}.}
\label{fig:sigtot}
\end{figure}

\section{Summary and Conclusions}

In the present paper we calculate the beam SSNA for small values
of $Q^2$ and provide physics arguments for the dominance of
contributions from collinear photons in the TPE mechanism. For
electron energies above the nucleon resonance region and small
$Q^2$ the contribution of collinear virtual photons leads to the
beam SSNA that is positive and has the order of
$m_e\sqrt{Q^2}\sigma^{\gamma p}_{tot}$, where $\sigma^{\gamma
p}_{tot}$ is the total photoproduction cross section on the
proton. This quantity is multiplied by the factor of the order
unity that includes a combination of double- and single-logarithm
terms. The fact that the beam SSNA does not decrease with the beam
energy at fixed $Q^2$ makes it attractive for experimental studies
at higher energies, for example, the energies to be reached at
Jefferson Lab after the forthcoming 12-GeV upgrade of CEBAF.

Since the collinear-photon-exchange effect follows from general properties of the rank-3 leptonic
tensor, it  should also take place at large values of $Q^2$ where the collinear kinematics has
to contribute with at least logarithmic enhancement.
The situation is different for unpolarized (or longitudinally polarized) electrons and
for the case of the normal beam polarization. In the first case the leptonic tensor is
proportional to the collinear photon 4--momentum $q_{1\beta}$ that leads to cancellation
of the collinear region contribution due to the condition of gauge invariance,
$q_{1\beta}H_{\mu\alpha\beta}
=0$, while this cancellation does not take place for the normal beam polarization.
Such behavior of the beam SSNA
does not depend on value of $Q^2$ and we verified this fact by considering excitations
of $N^*$ resonances in
the intermediate state. It means that the 3-momentum integration in Eq.(\ref{2})
in general produces logarithmic enhancement in the beam SSNA, unless the dynamics of the
non-forward Compton amplitude on a nucleon suppresses this contribution.

The beam SSNA is amplified by the effect similar to the Compton peak in
deep-inelastic scattering \cite{Spies} but with replacement of leptonic and hadronic blocks.
Namely, intermediate photon in the TPE box diagram can be collinear to the parent electron and
carry 4-momentum that is enough to create a large invariant mass of the intermediate
hadronic state. Moreover, the virtuality of this collinear photon is small and such
kinematics leads to a dynamical pole (and consequently enhancement) in the box diagram
with inelastic hadronic intermediate states. Emission of hard collinear photons is  known to enhance
helicity-flip effects, as was noted in the original article of  T.~D.~Lee and M.~Nauenberg \cite{Lee}
and recently discussed, for example, in the context of radiative muon decay \cite{radmuon}.

Because of the enhanced collinear-photon exchange contributions,
experiments measuring normal SSNA are sensitive to the
energy-weighted integrals of the same nucleon Compton amplitudes
(namely, their absorptive parts) that can be accessed in Compton
scattering experiments where at least one of the photons is real.
In contrast, calculations of TPE effects for unpolarized electron
scattering require knowledge of the nucleon Compton amplitude with
two space-like virtual photons. In relation to normal single-spin
asymmetries, we state that the TPE effects in experiments with
unpolarized (or longitudinally polarized) electrons, as opposed to
the normal polarized electron beams, probe different domains of
the non-forward nucleon Compton scattering which can be hidden in
the TPE amplitude with inelastic hadronic states. In the first
case, the entire 3-dimensional phase space in the TPE loop
integration contributes, while the regions of small photon
virtualities are suppressed. It justifies the `handbag' approach
with Generalized Parton Distributions for these observables, as
developed in Ref.\cite{Afanas}. For the second case, small
virtualities of the exchanged photons dominate the TPE integral.
In the kinematics $s>>-t$ above the resonance region, the beam
SSNA is defined by the total photoproduction cross section that,
in turn, is described by (soft) Pomeron exchange. Therefore the
soft diffractive (Pomeron) physics dominates the beam SSNA of
small-angle elastic electron-proton scattering associated with
{\it electron} helicity flip, in contrast to the known
helicity-conserving property of Pomeron exchange between {\it
hadrons}.

Large logarithms are also present in the QED radiative corrections
to a related observable, beam SSNA of polarized Moller scattering
\cite{Dixon}, where they are caused by initial- and final-state
radiation of collinear (real) photons.

When applying the approach \cite{Afanas} to the beam SSNA, as was done recently in Ref.\cite{Mark},
the contributions of hard collinear virtual photons are excluded in the `handbag' model
of the TPE interaction. Since the collinear photon region contributes with large
logarithmic enhancement, the beam SSNA is sensitive to `non-handbag' terms (for example,
Regge-exchange terms) in the TPE mechanism,
which are important to include in dynamical models of this observable.


\section*{Acknowledgements}

This work was supported by the US Department of Energy
under contract DE-AC05-84ER40150. N.M. acknowledges hospitality of Jefferson Lab,
where this work was completed. We thank I.~Akushevich, S.J.~Brodsky, E.~Beise, C.E.~Carlson,  T.W.~Donnelly, B.~Holstein,
K.~Kumar,  R.~Milner,  A.~Radyushkin, P.~Souder, M.~Vanderhaeghen, and S.P.~Wells for their interest to
this work and useful comments.

\appendix
\section{}
Taking into account Eqs. (\ref{2}) and (\ref{27}) one can write
the beam SSNA at small values of $Q^2$ as
\begin{equation}\label{A1}
A_n^e=\frac{m_e\sqrt{Q^2}\sigma^{\gamma p}_{tot}}{4\pi^3}\frac{F_1-\tau
F_2}{F_1^2+\tau F_2^2}
\int\frac{d^3k}{2E_k}\frac{(W^2-M^2)^2}{(s-M^2)^2}\frac{Q^2}{q_1^2
q_2^2}\ .
\end{equation}
The angular integration in Eq.(\ref{A1}) can be done by
introducing the Feynman parameter,
$$\int\frac{d\Omega_k}{q_1^2q_2^2}= \int\limits_0^1
dy\int\frac{d\Omega_k}{[-2m_e^2+2(kk_y)]^2}\ , $$
$$k_y=yk_1+(1-y)k_2 = (E_{k1}; y{\bf{k}_1} + (1-y){\bf{k}_2}) , $$
$$ (kk_y)=E_{k1}E_k-2k|{\bf{k}_y}|\cos{\theta_y}, \ \
d\Omega_k=d\Phi d\cos{\theta_y}\ . $$

Integration over $d\Omega_k$ and Feynman parameter $y$ is
straightforward, leading to
\begin{equation}\label{A.2}
\int\frac{d^3k}{2E_k}\frac{(W^2-M^2)^2}{(s-M^2)^2}\frac{Q^2}{q_1^2
q_2^2} = \frac{\pi}{2}\int_{m_{e}}^{E_{s}}\frac{dE_k}{k}(1-z)^2
L_1,
\end{equation}
 where $z=E_k/E_{k1}$, and
we extended the upper limit up
to $E_{k1} $ because the difference between the value of $E_k$ at
inelastic threshold (when $W^2=(m_{\pi}+M)^2$) and $E_{k1}$ is
negligible at large $s$, and the quantity $L_1$ is defined in Eq.(\ref{18}).

To calculate the integral in Eq.(\ref{A.2}), we note first that the region where
$k\simeq 0$ does not contribute because of the factor of $L_1$.
For this reason we can change integration with respect to
$E_k$ by integration over $k.$ Then we divide the integration
region into the following two parts, $ 0 < k < \lambda m_e$ and $\lambda m_e< k <
E_{k1}$, and choose the auxiliary parameter $\lambda$ in such a way
that
\begin{equation}\label{A3}
\lambda >> 1, \ \lambda m_e<<E_{k1} << \sqrt{Q^2}\lambda \ .
\end{equation}

In the first region we can neglect $E_k$ as compared with $E_{k1}$
and write the corresponding contribution in the form
$$I_1=\pi\int\limits_0^{\lambda m_e}\frac{dk}{k}L_1 =
\int\limits_0^{\lambda\sqrt{Q^2}/2E_{k1}} \frac{2\pi
dz}{\sqrt{1+z^2}}\log(z+\sqrt{1+z^2}) $$
\begin{equation}\label{A5}
=\pi\log^2\Big(\lambda\frac{\sqrt{Q^2}}{E_{k1}}\Big)\ .
\end{equation}
In the second region the quantity $\eta$ that enters $L_1$ is small and we have
\begin{equation}\label{A.4}
I_2=\pi\int\limits_{z_{\lambda}}^1\frac{dz}{z}(1-z)^2\Big(\log\frac{Q^2}{m_e^2}
+2\log\frac{z}{1-z}\Big)\ ,
\end{equation}
$$ z_{\lambda} = \frac{\lambda m_e}{E_{k1}} <<1\ .$$ The
integration in Eq.(\ref{A.4}) gives
$$I_2=\frac{\pi}{2}\Big(2\log\frac{1}{z_{\lambda}}\log\frac{Q^2}{m_e^2}
-2\log^2{z_{\lambda}} -3\log\frac{Q^2}{m_e^2}+\frac{2\pi^2}{3}
+2\Big) \ .$$

 In the sum
$I_1+I_2$ the auxiliary parameter $\lambda$ is cancelled and we
arrive at
\begin{equation}\label{A.6}
\int\frac{d^3k}{2E_k}\frac{(W^2-M^2)^2}{(s-M^2)^2}\frac{Q^2}{q_1^2 q_2^2} =
\end{equation}
$$\frac{\pi}{4}\Big(\log^2\frac{Q^2}{m_e^2}-6\log\frac{Q^2}{m_e^2}+\frac{4\pi^2}{3}+4\Big)\ . $$

\end{document}